\documentclass[preprint,amsmath,amssymb,prl]{revtex4-1}

\usepackage{amsmath}
\usepackage{amssymb}
\usepackage{graphicx}% Include figure files
\usepackage{dcolumn}% Align table columns on decimal point
\usepackage{bm}% bold math
\usepackage{graphics}
\usepackage{epsfig}
\usepackage{subfigure}
\usepackage{color}
\usepackage{enumitem}
\usepackage[tiny]{titlesec}

\newcommand\eq{\begin{equation}}
\newcommand\be{\begin{equation}}
\newcommand\eeq{\end{equation}}
\newcommand\ee{\end{equation}}
\newcommand\ar{\begin{eqnarray}}
\newcommand\ear{\end{eqnarray}}

\newcommand{\ii}{{\rm i}}

\newcommand*{\tdb}[1]{\overline{\overline{#1}}}

\newcommand{\nm}{\mbox{ nm}}
\newcommand{\dep}{\mbox{ $\Delta$}}

\newcommand{\epo}{\mbox{$\epsilon_\rho$}}

\newcommand{\epe}{\mbox{$\epsilon_e$}}

\newcommand{\cosine}{{\rm cos}}
\newcommand{\sine}{{\rm sin}}

\renewcommand{\Im}{\mbox{Im}}
\renewcommand{\Re}{\mbox{Re}}

\renewcommand{\deg}{^{\circ}}

\newcommand{\dv}{{\rm d}}

\begin{document}

\title{Cutoff-cladding waveguides: Subdiffraction guided modes near cutoff}

\author{C. H. Gan}
%\email{chg205@exeter.ac.uk}
\affiliation{ 
College of Engineering, Mathematics and Physical Sciences, University of Exeter, Exeter EX4 4QF, United Kingdom\\
}

%\ociscodes{(230.7370) Waveguides;~(260.1180) Crystal optics;~(310.6628) Subwavelength structures, nanostructures.}

%\doi{\url{http://dx.doi.org/10.1364/optica.XX.XXXXXX}}

\begin{abstract}
We propose a class of waveguides operating near cutoff such that electromagnetic energy is mainly bound to the cladding rather than the dielectric core to achieve subdiffraction confinement of light. The cladding incorporates an alternating stack of thin films that exhibit uniaxial form birefringence with a high contrast between the effective principal dielectric constants. In contrast to conventional dielectric waveguides, the effective modal length for the fundamental mode of the proposed waveguide diverges at a much slower rate for core thicknesses less than a twentieth of the illumination wavelength, and is comparable to that attainable for plasmonic waveguides. The proposed waveguide can be exploited for important applications such as optoelectronic integration, and the fundamental mode exhibits a near-uniform spatial field distribution potentially allowing for position-independent spontaneous emission enhancement effects.
\end{abstract}

\maketitle

%\section{Introduction}
Optical waveguides are the fundamental building blocks of integrated optoelectronic devices. As the demand for increased processing speed
and bandwidth continues to grow, high-capacity subwavelength channels are sought in place of traditional diffraction-limited waveguides to sustain 
nanoelectronic circuitry~\cite{bozhe_np_2010}. To this end, we propose a class of waveguides that we refer to as the \emph{cutoff-clad waveguide} (CCW), which operates near cutoff such that electromagnetic energy is mainly bound to the cladding rather than the dielectric core to achieve subdiffraction confinement of $p$-polarized light (transverse magnetic field). 
The cladding incorporates an alternating stack of thin films that exhibits uniaxial form birefringence~\cite{bornwolf} with a 
high contrast between the effective dielectric constants parallel and transverse to the optic axis. In contrast to 
conventional waveguides, the effective modal length for the fundamental mode of the proposed \emph{cladding waveguides} diverges at a much slower rate even for core 
thicknesses less than a twentieth of the wavelength, which renders them extremely attractive as integrated optical channels. 
As will be seen, the guided modes of the CCW can have effective modal lengths comparable to those for hybrid and channel plasmonic waveguides~\cite{oulton_njp_2008,gan_prx_2012}.
Although it is structurally similar to the all-dielectric layered waveguide system considered in Ref.~\cite{jacob}, we emphasize that in contrast, the CCW is operated near-cutoff with fields mainly confined to the cladding thus fully exploiting the subdiffraction modal confinement offered by the homogenized cladding. Additionally, we derive here the general condition for the existence of subdiffraction guided modes by considering the wave equations, and clarify that the condition for total internal reflection (TIR) in the core is satisfied for the bound modes so that it is not necessary to explain the waveguiding mechanism with the notion of partial-TIR introduced in Ref.~\cite{jacob}. Moreover, it is shown how absorption loss in the cladding (if present) can be mitigated. 
With their associated subdiffracted effective modal lengths, CCWs can potentially support position-independent spontaneous emission enhancement as predicted for 
single emitters in epsilon-near-zero channels~\cite{engheta_prl_2009,polman_prl_2013}.
This possibility stems from the virtually uniform spatial distribution across the core arising from the combination of the small core thickness and large transverse wavenumber near cutoff.
We also differentiate the layered cladding of the proposed CCW from plasmonic multilayers that exhibit indefinite or 
hyperbolic dispersion~\cite{smith_prl_2003,kivshar_np_2013}. The absence of localized surface plasmon polariton modes and high-wavevector components implies that the 
effective medium theory (EMT)~\cite{bornwolf} should provide a quantitatively accurate description of the homogenization of the 
layered cladding without the need to include nonlocal effects~\cite{elser_apl_2007,gan_ol_2010}.

\begin{figure}[ht!]
\centering
\includegraphics[width=3.0 in]{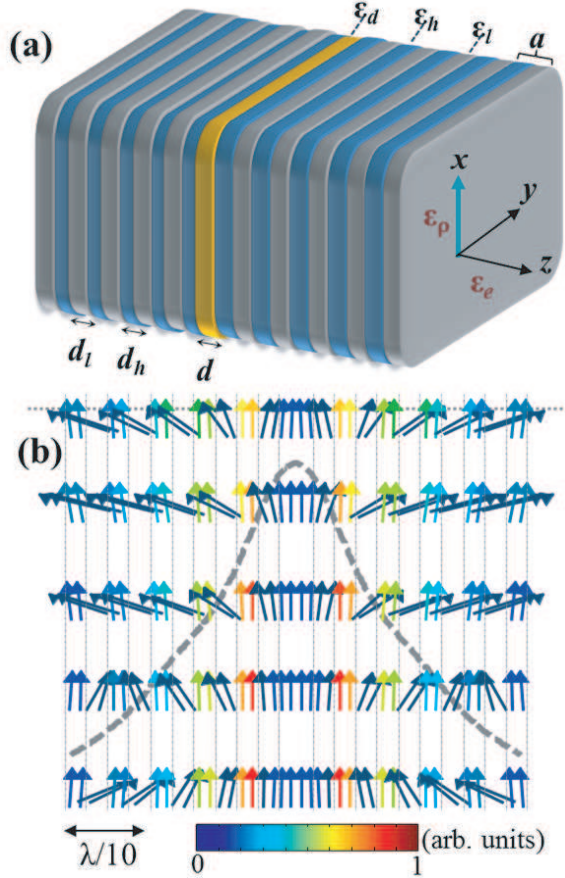} 
\caption{Geometry and waveguiding principle of the proposed cutoff-cladding waveguide (CCW). (a) Layout and main physical parameters. (b) Map of the time-averaged Poynting vector illustrating that electromagnetic energy is mainly carried in the cladding instead of the core. Vertical grey dotted lines indicate the boundaries between the different layers, and the overlying thick-dashed grey curve shows the transverse magnetic field $H_y$ associated with the fundamental mode. The waveguide is bounded by an air interface near the top of the map as shown by the dotted grey horizontal line.%, in which it has been taken that $d = 0.02\lambda$, $a = 40 \nm$, $n_{d} = 4.3, n_h = 5.5, n_l = 1.1$, and $\lambda = 1550 \nm$. 
 }
\vspace*{-0.15in} 
\end{figure}

A schematic of the CCW, which consists of a narrow dielectric core ($n_d$) of thickness $d$ cladded with alternating stacks of thin films of high $(n_h)$ and low ($n_l$) refractive index, is shown in Fig.~1(a). The multilayered cladding gives rise to form birefringence with effective dielectric constants $\epo = f\epsilon_{h} + (1-f)\epsilon_{l}$ and $\epe = \epsilon_{h} \epsilon_{l} / [f\epsilon_{l} + (1-f)\epsilon_{h}]$, parallel and perpendicular to the plates, respectively. Here $\epsilon_{h, l} = n_{h, l}^2$, the fill factor $f = d_h/a$, where $a = d_h + d_l$ and $d_h\,(d_l)$ is the thickness of the high (low) index layer. 
For a bound mode with effective index $n$ propagating along the $x-axis$, the wave equation for the magnetic field $H_y$ in the homogenized uniaxial cladding as obtained with the EMT is (see supplementary information {\bf S1} for details)
\eq\label{hywave}
\frac {\partial^2 H_y}{\partial z^2} + \left[\frac{\epo}{\epe}(\epe - n^2)k_0^2 \right] H_y = 0 \,,
\eeq
where $k_0$ is the free-space wavenumber. The above wave equation applies equally well for any individual layer in the waveguide if $\epo \rightarrow  \epe \rightarrow  \epsilon_{q}\,(q = d, h, l)$. 
For nontrivial solutions of $H_y$, the transverse wavenumber in the homogenized cladding and core according to~Eq.~\eqref{hywave} is respectively,
\eq\label{guide_con1} 
\alpha_c^2 =  \frac{\epo}{\epe} (n^2 - \epe) k_0^2,  \quad \alpha_d^2 =  (n^2 - \epsilon_{d}) k_0^2 \,,
\eeq
where $\frac {\partial}{\partial z} \rightarrow \ii k_z$ and $k_{z}^{(c, d)}~=~\ii \alpha_{c, d}$. The sign of the term in parentheses in~Eq.~\eqref{guide_con1} determines the wave character in each section of the waveguide, with the negative (positive) sign corresponding to oscillating (exponentially decaying) fields. Thus, one may deduce from~Eq.~\eqref{guide_con1} that bound modes are supported provided
\eq\label{ccw_con}
\epsilon_d > n^2 > \epe
\eeq 
and that the subdiffraction confinement is achieved for $\epo >> \epe$ as the field intensity decay length in the cladding layers is proportional to $\epe/\epo$. 
The reflection coefficient between the core-cladding interface is $\Gamma = (\epo \alpha_d - \epsilon_d \alpha_c)/(\epo \alpha_d + \epsilon_d \alpha_c)$, which together with~Eq.~\eqref{guide_con1} shows that the condition for TIR in the core 
is $\epsilon_d~>~\epe$~(see supplementary information {\bf S1}) in accordance with the condition~(\ref{ccw_con}). Hence, the TIR condition is always satisfied for the bound modes. 
For $\epsilon_h > 2 \epsilon_l > 0$, it is straightforward to show that the ratio $\epo/\epe$ is maximal for fill factor $f = 0.5$, hence it is taken that $d_h = d_l$ throughout our analysis. 
Since $n^2 \rightarrow \epe$ at cutoff, one may express the cutoff condition for the $m^{\rm th}-$order mode as $\frac{d}{\lambda} \lesssim \frac{m}{2\sqrt{\epsilon_d-\epe}},$
where $\lambda = 2\pi/k_0$ is the free-space wavelength, $m = \{0, 1, 2,...\}$, and the fundamental $m = 0$ mode has no cut-off as expected. The cut-off condition shows that for sufficiently small $\epe$, well-confined higher 
order modes ($m \lesssim n_d$) can be excited with subwavelength core thicknesses, a feature that is unique to CCWs. Another distinguishing feature of the CCW is that the energy is mainly confined in the uniaxial cladding, as shown in the map of the time-averaged Poynting vector for a cladding waveguide with core thickness $d = 0.02\lambda$ in Fig.~1(b). The concentration of the energy in the low-index rather than the high-index layer is due to the discontinuity of the $z-$component of the electric field $E_z$ at the interfaces (i.e. $E_z^{(h)}/E_z^{(d)} = \epsilon_{d} /\epsilon_{h}$ and $E_z^{(l)}/E_z^{(h)} = \epsilon_{h} /\epsilon_{l}$). For all simulations, which have been computed with the aperiodic-Fourier Modal Method (a-FMM)~\cite{caojosa},
it is taken that $n_{d} = 4.3, n_h = 5.5, n_l = 1.1$, the 
period $a = 40 \nm$, and $\lambda = 1550 \nm$ unless otherwise specified. Predictions of the EMT are also presented where it is prudent to make comparisons. In the calculations, the field components are normalized such that the guided mode has unity Poynting flux.  Archetypal materials for the core and high-index layers at $\lambda = 1550 \nm$ are germanium~\cite{potter_pr_1966} and crystalline 
germanium telluride (GeTe)~\cite{bahl_jap_1969} respectively, and the film thicknesses are within the limits of current nanofabrication techniques. Low-index thin films are more challenging to synthesize, but let us note that sub-micron thin films of porous silica with refractive index $\sim~1.1$ in the visible to near-infrared have been reported~\cite{jain_tsf_2001,xi_np_2007}. Alternatively, phonon-polariton materials such as lithium fluoride can exhibit a permittivity close to zero with small absorption loss in the THz regime~\cite{soukoulis_prb_2013}.  

\begin{figure*}[ht!]
%\vspace*{-0.2in} 
\centering
\includegraphics[width=6.8 in]{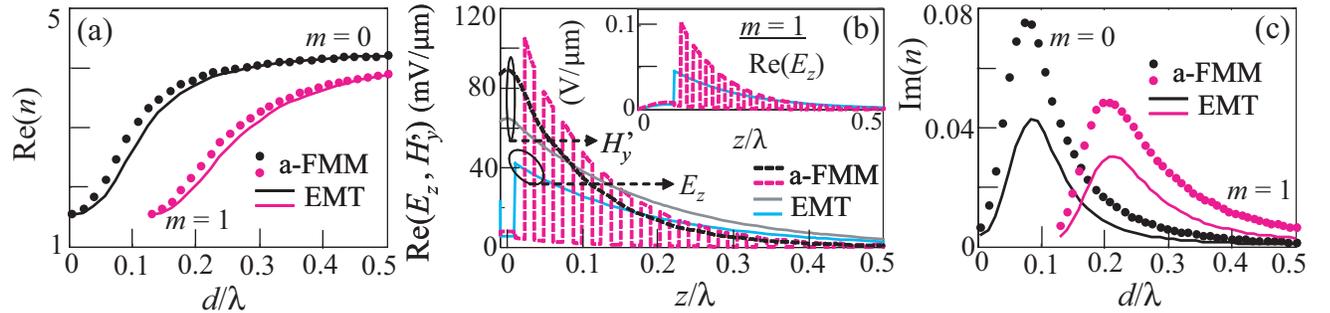} 
\vspace*{-0.1in} 
\caption{Dispersion and modal profile for the lowest order even and odd modes ($m = 0, 1$) of a CCW. (a) Effective index in the absence of absorption. (b) Transverse field components $H_y' = \sqrt{\mu_0/\epsilon_0}\, H_y$ and $E_z$ of the fundamental mode for $d = 0.02\lambda$. Inset: $E_z$ for the $m = 1$ mode. (c) Imaginary part of the effective index taking into account material loss in the GeTe layers.%~($n_h = 5.5 + \ii 0.3$).
}
\vspace*{-0.15in} 
\end{figure*}

The effective index $n$ of the the lowest order even ($m = 0$) and odd ($m = 1$) modes have been computed with the a-FMM and the EMT, and both methods yield results in excellent agreement %especially for $d/\lambda$ close to cutoff and for $d/\lambda \gtrsim 0.4$ 
as shown in Fig.~2(a). Let us note that the EMT predictions indeed approach those of the a-FMM for $a/\lambda \rightarrow 0$.
As predicted by the EMT, the cut-off for the $m = 1$ mode occurs close to $d = \lambda/8$, and the effective index $n \rightarrow \sqrt{\epe}$ as the modes approach cutoff or $d \rightarrow 0$.
The transverse magnetic field ($H_y$) and electric field ($E_z$) components for the fundamental mode for $d = 0.02\lambda$ are shown in Fig.~2(b). For $H_y$, which is continuous across the CCW, there is good agreement between the a-FMM (black dashed curve) and EMT (grey solid curve) calculations. For $E_z$, even though the EMT produces an averaged response (solid blue curve) due to the homogenization, it is seen that its prediction of the field decay length matches closely that of the exact calculation (red dashed curve). In the core region ($z \leq |0.01 \lambda|$), it is seen that $E_z$ is spatially uniform across and $H_y$ varies marginally as a result of the large transverse wavenumber arising from the small effective index $n$ (see~Eq.~\eqref{guide_con1}). As $d \rightarrow 0$, the longitudinal electric field component $E_x$ of the fundamental mode ($m = 0$) is small everywhere along the cross section of the CCW as a result of the slowly-varying $H_y$ (see supplementary information {\bf S1}), hence $E_z$ is the dominant component. Thus position-independent emission enhancement of $z$-polarized dipole emitters embedded in the core can in principle be realized in a fashion similar to that predicted for molecules in epsilon-near-zero channels~\cite{engheta_prl_2009}. 
The $E_z$ component for the odd ($m = 1$) mode for $d = 0.15\lambda$ is shown in the inset of Fig.~2(b). 
Unlike the fundamental 
mode, the magnitude of the $E_x$ component (not shown) can be larger or comparable to $|E_z|$ inside the core due to the zero-crossing of $H_y$ at $z = 0$. %In comparison to the $m = 0$ mode, the core thickness is much larger due to the cutoff condition and the spatial variation of the fields are dramatically increased. 
But it is worth noting that the physical thickness in both 
the two cases ($m = 0, 1$) can be less than the free-space wavelength $\lambda$, and that about 10-20 periods of the cladding layers are sufficient to confine the guided modes. 
Before we quantify the confinement capability of the CCW, let us briefly account for the influence on the waveguide dispersion in the presence of absorption loss. Material loss in the core would only contribute marginally since most of the electromagnetic energy is carried in the cladding near cutoff. Loss in the high-index and low-index layers should be limited so that $\Re(\epsilon_{h,\,l}) >> \Im(\epsilon_{h,\,l})$ for reasonable propagation lengths of the guided modes. 
For instance, taking into account the absorption loss present in the GeTe layers ($n_h = 5.5 + \ii 0.3$) in the above CCW, the imaginary part of the effective index has been calculated and is shown in Fig.~2(c). 
For thick cores $d/\lambda \gtrsim 0.5$, most of the mode energy is confined in the core and dissipation arising from cladding loss is negligible. As $d$ is decreased, more energy is transferred into the cladding layers and the absorption loss increases, peaking at $d = 0.075\lambda~(0.2 \lambda)$ for the fundamental even (odd) mode. 
As $d$ is further reduced, the response of the CCW becomes dominated by the homogenized claddings as  $n \rightarrow \sqrt{\epe}$. %The absorption loss for extremely thin cores therefore depends on $\epe$ rather than the material loss of individual layers. %For this specific example, $\sqrt{\epe} = 1.53 + \ii 3.18 \times 10^{-3}$; with an imaginary part about 100 times smaller than GeTe.
This explains the rapid decrease of the absorption of the CCW observed in Fig.~2(c) as the core thickness approaches cutoff. %Similar trends are observed for other combinations of lossy materials for the cladding layers as long as the CCW remains weakly absorbing. 
Let us remark that material loss in the low-index layers is more likely to result in stronger absorption effects than loss in the high-index layers because (i) the $E_z$ component is much larger in the low-index layers, and (ii) the ratio $\Re(\epsilon_{l})/ \Im(\epsilon_{l})$ remains large only for very small values of $\Im(\epsilon_{l})$.
Our calculations also show that the presence of a modest amount of absorption does not lead to any visible difference in the waveguiding characteristics of the CCW such as the fraction of power in the cladding, the real part of the effective index (see Fig.~2(a)), and the effective modal length which is discussed below.

\begin{figure}[ht!]
\centering
\includegraphics[width=2.2 in]{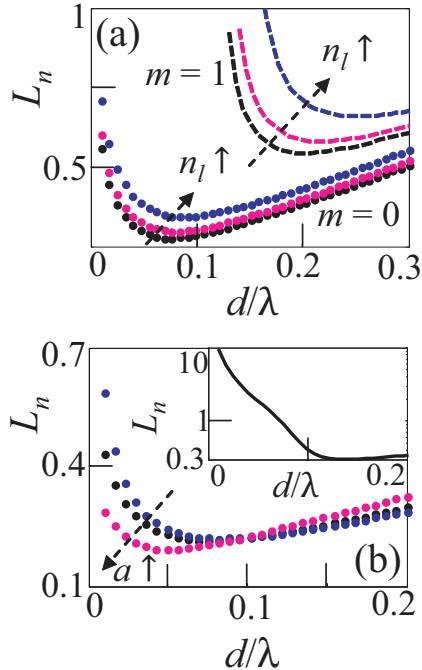} 
\caption{Subdiffraction confinement of guided modes of the CCW. (a) Normalized effective modal length $L_n$ for the lowest order even and odd modes for $n_l = \{1.1, 1.5, 2.1\}$. (b) $L_n$ for the fundamental mode for $a = \{20, 40, 80\} \nm$. Inset: $L_n$ for a conventional dielectric waveguide with core and cladding index taken to be 4.3 and 1.5, respectively. Also note the logarithmic scale for the vertical axis.}
\vspace*{-0.15in} 
\end{figure}

To demonstrate subdiffraction confinement of the CCW, we have computed the effective modal length $L_m$ of the fundamental even and odd modes. To compare to plasmonic waveguides and to remain conservative in our estimate of $L_m$, we adopt 
the measure~\cite{oulton_njp_2008}
$L_m = \left[ \int W(z) \dv z \right]^2 / \int W(z)^2 \dv z$, where $W(z) = \frac{1}{4} \left[ \epsilon_q \epsilon_0 |E|^2 + \mu_0 |H|^2 \right]$ is the time-averaged 
electromagnetic energy density~\cite{ruppin_pla_2002} in the CCW. Figure~3(a) shows the normalized effective modal length $L_n = 2L_m/\lambda$ with $n_l = \{1.1, 1.5, 2.1\}$ for the fundamental even and odd modes ($m = 0,\,1$). Both modes offer subdiffraction confinement comparable to hybrid plasmon-polariton and cylindrical metallic 
waveguides, with $L_m < \lambda/2$ for core thicknesses $d$ as small as $0.01\lambda$ for the even mode. It is remarkable that such a tight confinement can be attained without subdiffracted plasmonic modes. 
As indicated by~Eq.~\eqref{guide_con1}, $L_n$  decreases for decreasing $n_l$ for a given index $n_h$ (taken to be GeTe) of the high-index layers. For the even mode, the effective modal length decreases steadily for large core thicknesses to a minimum of $\sim 0.15 \lambda \,(L_n \approx 0.3)$ at $d \sim 0.05\lambda$, 
after which it turns around and increases with a slope that gradually diverges as $d \rightarrow 0$. 
As the EMT tends to underestimate the effective index $n$ (see Fig.~2(a) and Fig.~2(c) for instance),  
the rate at which $L_n$ diverges can be controlled to some extent by increasing the period $a$. This is clearly depicted in Fig.~3(b), where it is seen that $L_n$ diverges more slowly after the turning point if the period $a$ is doubled to $80 \nm$ while decreasing $a$ has the reverse effect. 
However absorption loss, if present, would also increase with the period $a$.
Also, if the period is not sufficiently small~\cite{gan_ol_2010} such that $a/\lambda \lesssim 0.1$, the layered system no longer exhibits form birefringence and the EMT does not apply. While such structures with non-birefringent cladding may still support guided modes, their behavior is beyond the scope of the present study. Finally, it is 
worth noting that in comparison to conventional waveguides (see inset of Fig.~3(b)), $L_n$ for the fundamental mode of the CCW is about an order smaller for $d/\lambda \lesssim 0.05$, asserting its capability to support subdiffracted guided modes.

In summary, we have demonstrated theoretically the proposed class of cutoff-cladding waveguides (CCWs) exploits the subdiffraction confinement offered by homogenized claddings to attain effective modal lengths comparable to plasmonic waveguides. Our findings reveal that the CCW poses a potential design for optoelectronic integration and for supporting phenomenal spontaneous emission enhancement effects in cavity quantum electrodynamics.

I acknowledge G. R. Nash for useful discussion.

\newpage

\begin{center}{\bf Supplementary information S1 -- Cutoff-cladding waveguides: Subdiffraction guided modes near cutoff}
\end{center}
\begin{center}{Choon How Gan}\end{center}

%\affiliation{Laboratoire Charles Fabry de l'Institut d'Optique, CNRS,\\
%Univ Paris-Sud, Campus Polytechnique, 91127 Palaiseau Cedex, France}
\begin{center}{College of Engineering, Mathematics and Physical Sciences, University of Exeter,\\ Exeter EX4 4QF, United Kingdom}
\end{center}

\renewcommand{\theequation}{S1-\arabic{equation}}
 % redefine the command that creates the equation no.
\setcounter{equation}{0}  % reset counter 
\section{A. Derivation of the general condition for the existence of subdiffraction guided modes in a dielectric waveguide cladded with uniaxial media}
Here, the general condition for the existence of subdiffraction guided modes in the proposed cutoff-cladding waveguides (CCWs) are derived based on the set of wave equations that governs wave propagation in the system of materials. It is taken that the constituent materials have real positive dielectric constants with a small imaginary part if material absorption is present, and that they are non-magnetic. 
The EMT~[1] is used to describe the homogenization of the cladding layers, which collectively form an uniaxial medium with the optic axis directed along $z$~(see Fig.~1(a) of main text). However for generality, we allow for the direction of the optic axis of the anisotropic cladding to be either parallel or perpendicular to the material interfaces in the following derivation. 
The dielectric tensor of the homogenized cladding layers then takes on the form
\eq\label{emten_b0}
\tdb{\epsilon}\,\, =\,\,
\begin{bmatrix} 
\epsilon_\rho + \sine^2\vartheta \dep & 0 & 0\\
0 & \epsilon_\rho & 0\\
0 & 0 & \epsilon_\rho + \cosine^2\vartheta \dep\\
\end{bmatrix} 
\,\,,\\
\eeq
where $\dep = \epe - \epo$, $\epsilon_\rho$ and $\epsilon_e$ are the effective ordinary and extraordinary dielectric constants, and $\vartheta = 0$ ($\vartheta = \pi/2$) corresponds to the optic axis directed along the $z-$axis ($x-$axis).

From Maxwell equations in the absence of free charges,${\bf \nabla} \times {\bf E}  + \frac {\partial {\bf B}}{\partial t}  = 0, {\bf \nabla} \times {\bf H} - \frac {\partial {\bf D}}{\partial t} =  0$,
%\eq\label{grpcsi3}
%{\bf \nabla} \times {\bf E}  + \frac {\partial {\bf B}}{\partial t}  = 0, \quad {\bf \nabla} \times {\bf H} - \frac {\partial {\bf D}}{\partial t} =  0  \,\,,
%\eeq
we have for $p-$polarized modes propagating along $x$ 
\begin{subequations}\label{ehmax}
\begin{align}
\ii \beta E_z - \frac {\partial E_x}{\partial z} & = \ii \omega \mu_0 H_y \,,\label{hmax2b}\\
\frac {\partial H_y}{\partial z} & = -\ii \omega \epsilon_0 \epsilon_{xx} E_x \,, \label{emax1a}\\
\beta H_y &  = \omega \epsilon_0 \epsilon_{zz} E_z \,, \label{emax1c}
\end{align}
\end{subequations} 
where $\beta = n k_0$ is the propagating constant, and
\eq\label{epxxzz}
\begin{bmatrix}      
\epsilon_{xx}   \\ 
\epsilon_{zz} 
\end{bmatrix}
\,\,=
\begin{bmatrix} 
\cosine^2\vartheta & \sine^2\vartheta \\
\sine^2\vartheta & \cosine^2\vartheta 
\end{bmatrix} 
\,\,
\begin{bmatrix}      
\epo \\ 
\epe 
\end{bmatrix}
\,\,, \quad \textrm {($\vartheta = 0$ or $\pi/2$)}\,\,.
\eeq 
Taking the partial derivative $\frac {\partial}{\partial z}$ of~\eqref{emax1a} and eliminating the electric field components then yields the wave equation
\eq\label{shywave}
\frac {\partial^2 H_y}{\partial z^2} + \left[\frac{\epsilon_{xx}}{\epsilon_{zz}}(\epsilon_{zz} k_0^2 - \beta^2) \right] H_y = 0 \,,
\eeq
which is identical to~Eq.~(1) of the main text for $\vartheta = 0$. As explained in the main text, the CCW will therefore support guided modes provided the condition $\epsilon_d > n^2 > \epsilon_{zz}$ or
\begin{equation}\label{guide_con}
\epsilon_d > \left \{ \begin{array}{ll}
\epe\,\, & \textrm {\quad \quad  ($\vartheta = 0$) }\\
\epo \, \, & \textrm {\quad \quad ($\vartheta = \pi/2$) }
\end{array} \right.
\end{equation}  
is satisfied. We note that the above condition~(\ref{guide_con}) is independent of $\epo\,(\epe)$ for the case the optic-axis is oriented along $z\,(x)$. 
%~\eqref{hywave} of the main text. 

\section{B. Demostrating that the condition for total internal reflection in the core is satisfied for bound modes of the CCW}
Next, we show that within the EMT,
the condition for total internal reflection (TIR) to occur is also satisfied for bound modes of the CCW. This is identical to the case of conventional isotropic dielectric waveguides, where bound modes are supported only for cores with a higher refractive index than the claddings. 
Let us focus on the case with the optic axis of the homogenized cladding layers aligned along the $z-$axis ($\vartheta = 0$), since results for the other orientation ($\vartheta = \pi/2$) can be obtained by interchanging $\epo \leftrightarrow \epe$. 
The condition for TIR to occur at a dielectric-anisotropic medium interface can be inferred
from the relevant Fresnel coefficients.
It is straightforward to show from~\eqref{ehmax} that the field components in the uniaxial cladding can be expressed in 
terms of $H_y$ as $\Psi = [E_x,\, H_y,\, E_z] = \frac{1}{\omega \epsilon_0}[-\ii \alpha_c/\epo, 1, \beta/\epe] H_y$. 
Without loss of generality, let us consider a planar semi-infinite core-uniaxial cladding interface at $z = 0$ to calculate the Fresnel coefficients. The tangential field components in each medium can be expressed as
\begin{equation}\label{B_hy}
H_y =  \exp{(\ii \beta x)} \left \{ \begin{array}{ll}
\exp{(-\alpha_{d} \,z)} + \Gamma\,\exp{(\alpha_{d} \,z)}  \, & \textrm {$z < 0$ }\\
\tau \,\exp{(-\alpha_c \,z)}  \, & \textrm {$z > 0$ }
\end{array} \right.
\end{equation} 
and
\begin{equation}\label{B_ex}
E_x =  -\frac{\exp{(\ii \beta x)}}{\omega \epsilon_0} \left \{ \begin{array}{ll}
\frac{\ii \alpha_d}{\epsilon_d}\left[\, \,\exp{(-\alpha_{d} \,z)} - \Gamma \,\exp{(\alpha_{d}\,z)} \,\right ] \, & \textrm {$z < 0$ }\\
\dfrac{\ii \alpha_c}{\epo} \tau \,\exp{(-\alpha_c \,z)}  \, & \textrm {$z > 0$ }
\end{array} \right.
\end{equation} 
where $\Gamma$ and $\tau$ are reflection and transmission coefficients defined for $H_y$, and $\alpha_{c,\,d}$ are 
as defined in~Eq.~(2) of the main text.
Matching boundary conditions at $z = 0$ yields 
\begin{equation}\label{B_r1_xz}
\Gamma = \dfrac{\epo \alpha_d - \epsilon_d \alpha_c}{\epo \alpha_d + \epsilon_d \alpha_c}\,\,,\quad \tau = \dfrac{2\epo \alpha_d}{\epo \alpha_d + \epsilon_d \alpha_c} \,\,.
\end{equation} 
Neglecting absorption, $\alpha_d$ is pure imaginary for bound modes of the CCW (see~Eq.~(2) main text), which also follows by considering $\beta^2 = \epsilon_d k_0^2 \sine^2\theta   \leq \epsilon_d k_0^2$, where $0 \leq \theta \leq 90\deg$ is the angle of incidence with respect to the $z-$axis.
Total internal reflection occurs 
provided $\alpha_c$ is real such that $|\Gamma| = 1$, which requires that $\epsilon_d > \epsilon_e$, corresponding to the condition~(\ref{guide_con}) for the CCW to support guided modes. Thus we see that the condition for TIR to occur is satisfied for the bound modes, and it is not necessary to consider the {\em relaxed}-TIR mechanism
proposed in Ref.~[2].

\section{References}
\begin{enumerate}[label={[\arabic*]}]
\item M. Born and E. Wolf, {\em Principles of Optics}, 7th ed., Cambridge University Press, Cambridge, 2002.
\item S. Jahani, and Z. Jacob, ``Transparent subdiffraction optics: nanoscale light confinement without metal,'' Optica \textbf{1},   96--100 (2014).
 
%\bibitem{bornwolf}M. Born and E. Wolf, {\em Principles of Optics}, 7th ed., Cambridge University Press, Cambridge, 2002.

%\bibitem{jacob}S. Jahani, and Z. Jacob, ``Transparent subdiffraction optics: nanoscale light confinement without metal,'' Optica \textbf{1},   96--100 (2014).

\end{enumerate}

\end{document}